\documentclass[prb,showpacs,twocolumn,amssymb,amsmath]{revtex4-1}

\usepackage{color}
\usepackage{bm}
\usepackage{graphicx}
\usepackage[english]{babel}

\newcommand{\be}{\begin{equation}}
\newcommand{\ee}{\end{equation}}
\newcommand{\bea}{\begin{eqnarray}}
\newcommand{\beaa}{\begin{eqnarray*}}
\newcommand{\eea}{\end{eqnarray}}
\newcommand{\eeaa}{\end{eqnarray*}}

\begin{document}

\title{\bf\Large {Phase diagrams of $\rm La_{1-x}Ca_xMnO_3$ in Double Exchange Model with added antiferromagnetic and Jahn-Teller interaction}}

\author{Vasil Michev}
\author{Naoum Karchev}\email{naoum@phys.uni-sofia.bg}

\affiliation{Department of Physics, University of Sofia, 1164 Sofia, Bulgaria}

\begin{abstract}
The phase diagram of the multivalent manganites $\rm La_{1-x}Ca_xMnO_3$, in space of temperature and doping $x$, is a challenge for the theoretical physics. It is an important test for the model used to study these compounds and the method of calculation. To obtain theoretically this diagram for $x<0.5$, we consider the two-band Double Exchange Model for manganites with added Jahn-Teller coupling and antiferromagnetic Heisenberg term. In order to calculate Curie and N\'{e}el temperatures we derive an effective Heisenberg model for a vector which describes the local orientation of the total magnetization of the system. The exchange constants of this model are different for different space directions and depend on the density of $e_g$ electrons, antiferromagnetic constants and the Jahn-Teller energy. To reproduce the well known phase transitions from A-type antiferromagnetism to ferromagnetism at low $x$ and C-type antiferromagnetism to G-type antiferromagnetism at large $x$, we argue that the antiferromagnetic exchange constants should  depend on the lattice direction. We show that ferromagnetic to A-type antiferromagnetic transition results from the Jahn-Teller distortion. Accounting adequately for the magnon-magnon interaction, Curie and N\'{e}el temperatures are calculated. The results are in very good agreement with the experiment and provide values for the model parameters, which best describe the behavior of the critical temperature for $x<0.5$.
\end{abstract}

\pacs{75.47.Lx, 63.20.kd, 71.27.+a, 75.30.Ds}

\maketitle

\section{Introduction}

Manganites remain one of the most studied classes of materials in modern condensed matter physics, with many unanswered questions to keep us interested in them. At the same time, their diverse properties still hold the promise of many potential applications, and while low temperatures and high magnetic fields required may limit their availability in consumer electronics, they still provide excellent opportunities for advancing nanotechnology. This is why in a series of papers\cite{Us1,Us2,Us3} during the past few years we have examined different aspects of the physics of manganites and constructed and tested a model which correctly describes their magnetic properties, such as the observed phases and the Curie temperature. In the present paper we will apply this model to a ``real-world'' compound, such as ${\rm La_{1-x}Ca_xMnO_3}$, and we will try to determine the values of the model parameters which best correspond to the experimental observations for the Curie temperature.

\begin{figure*}[!ht]
\includegraphics[width=.7\linewidth]{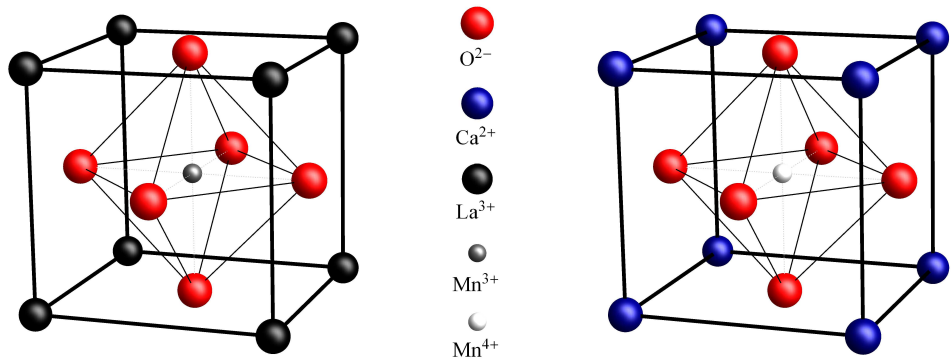}
\caption{(Color online) Unit cell of a perovskite, shown for $\rm LaMnO_3$ (left) and $\rm CaMnO_3$ (right).}\label{fig1}
\end{figure*}

Mixed-valence manganites with perovskite structure were first described by Jonker and Van Santen\cite{Jonker1,Jonker2,Jonker3} in series of papers in the 1950s. These materials can be regarded as solid solutions between end members such as ${\rm LaMnO_3}$ and ${\rm CaMnO_3}$, leading to mixed-valence compounds such as ${\rm La_{1-x}Ca_xMnO_3}$. The general chemical formula for a perovskite compound is ${\rm ABX_3}$, where ${\rm A}$ and ${\rm B}$ are two cations of different sizes and ${\rm X}$ is an anion that bound to both. In the case of mixed-valence manganites, we have two different ${\rm A}$ type ions, usually alkaline metal or lanthanoid/rare-earth ions such as ${\rm Ca^{2+}}$, ${\rm Sr^{2+}}$, ${\rm La^{3+}}$. ${\rm B}$ type ions are the smaller ${\rm Mn^{3+}}$ or ${\rm Mn^{4+}}$ ions and ${\rm X}$ is oxygen. The ideal cubic-symmetry structure has the manganese cation in 6-fold coordination, surrounded by an octahedron of oxygen anions, and the ${\rm A}$ cation in 12-fold cuboctahedral coordination. The perovskite structure is shown in Figure \ref{fig1} for both end compounds.

\begin{figure*}[!ht]
\includegraphics[width=.2\linewidth]{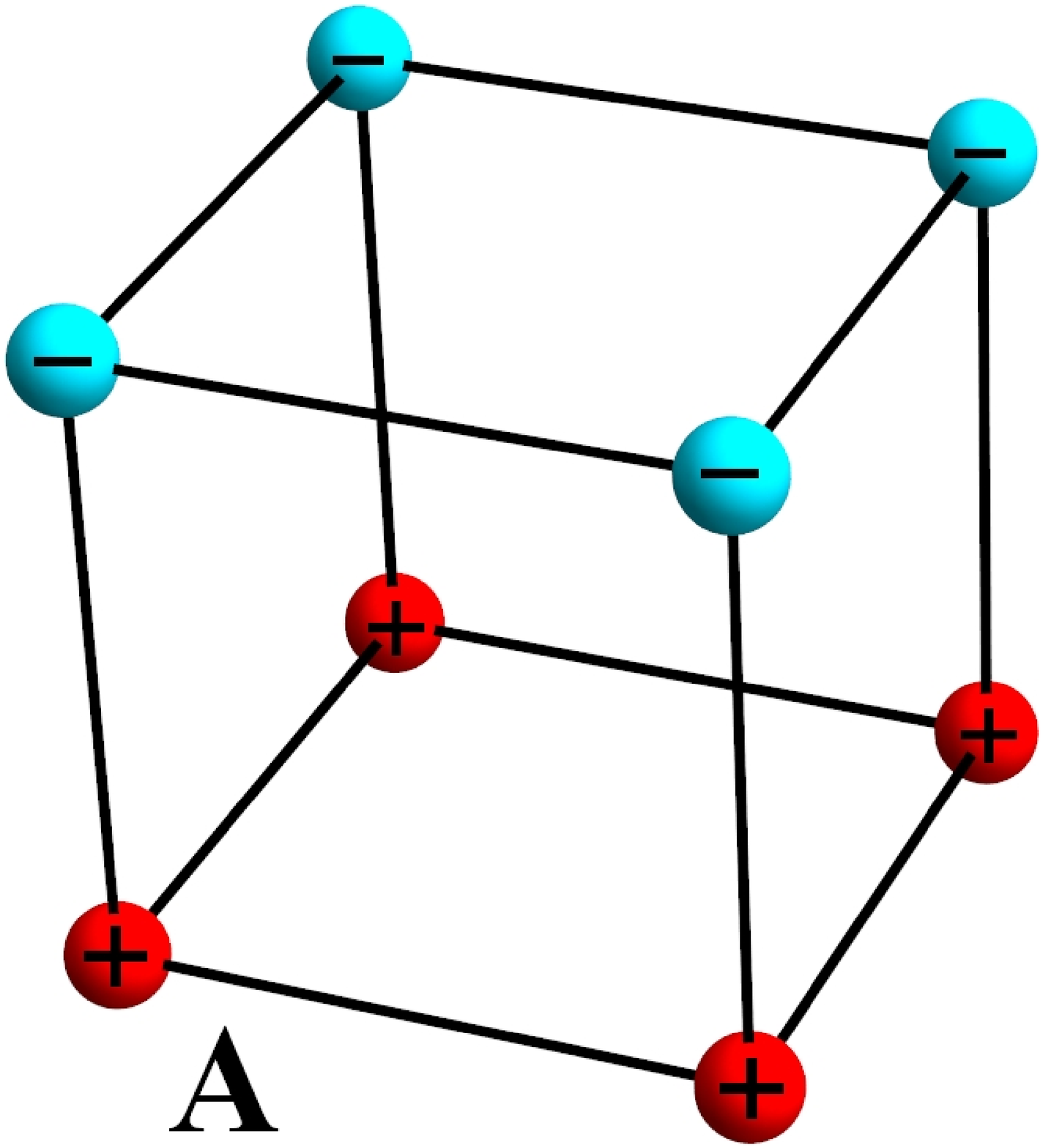}\hskip 1cm \includegraphics[width=.2\linewidth]{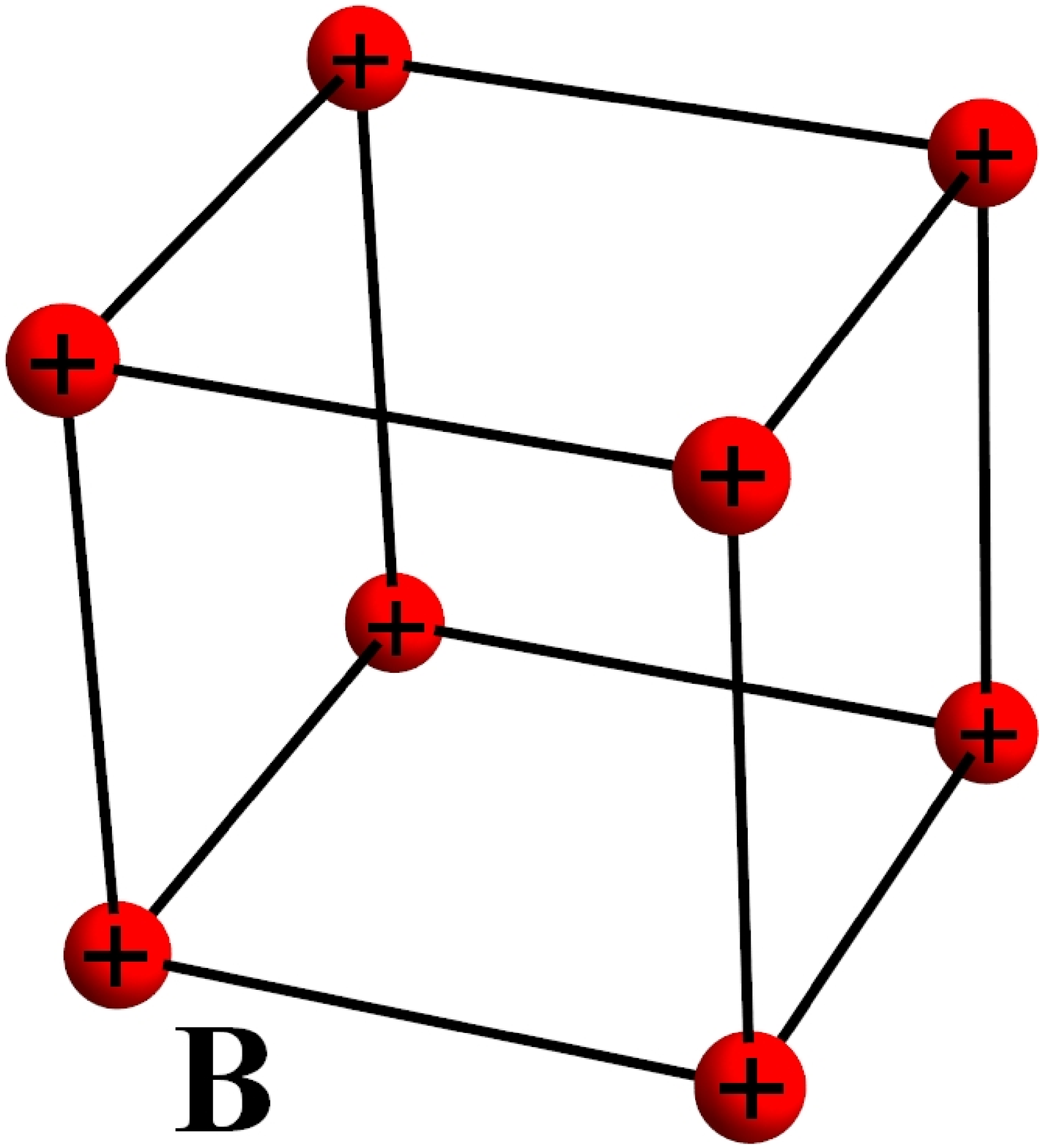} \hskip 1cm \includegraphics[width=.2\linewidth]{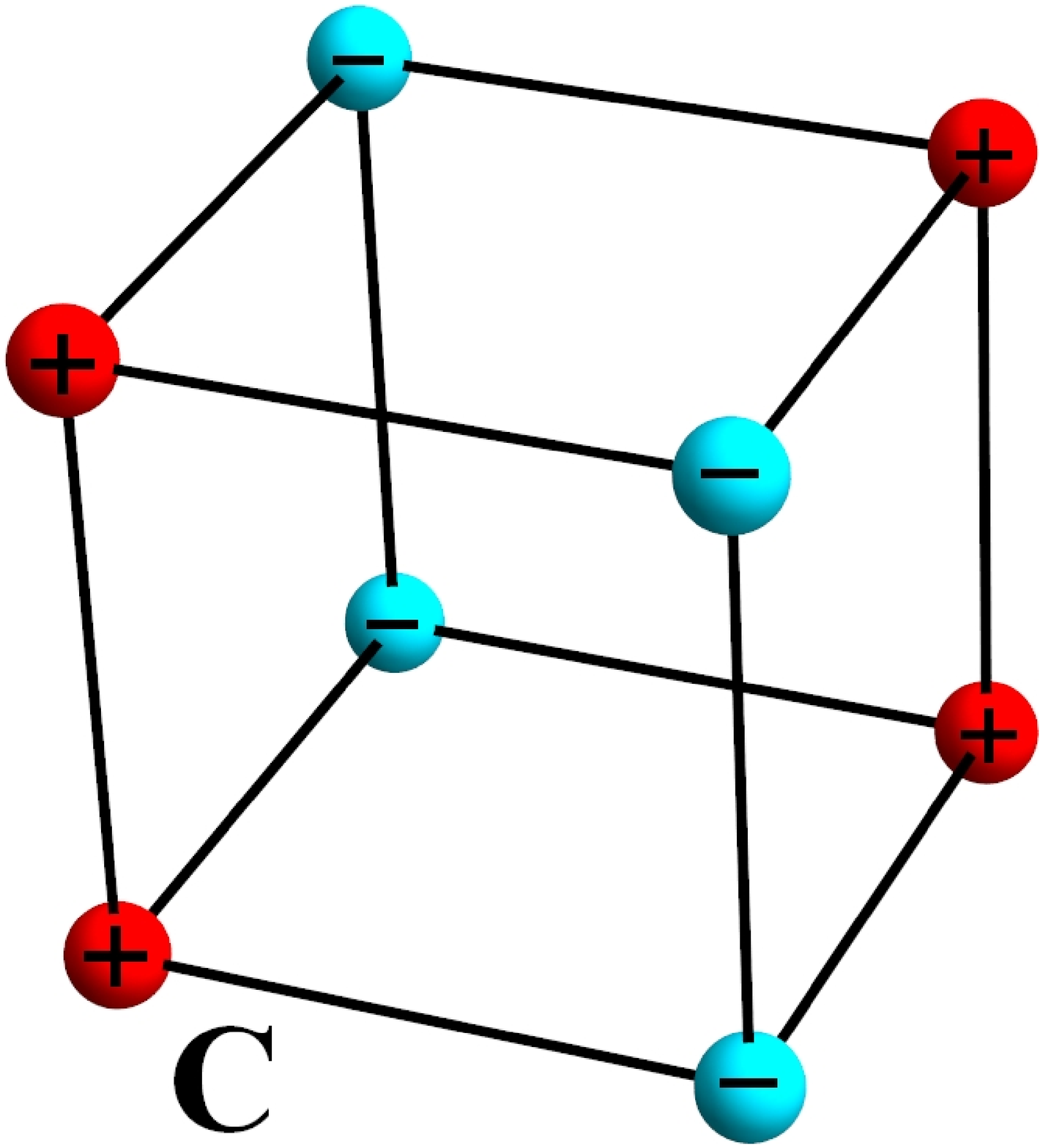} \hskip 1 cm \includegraphics[width=.2\linewidth]{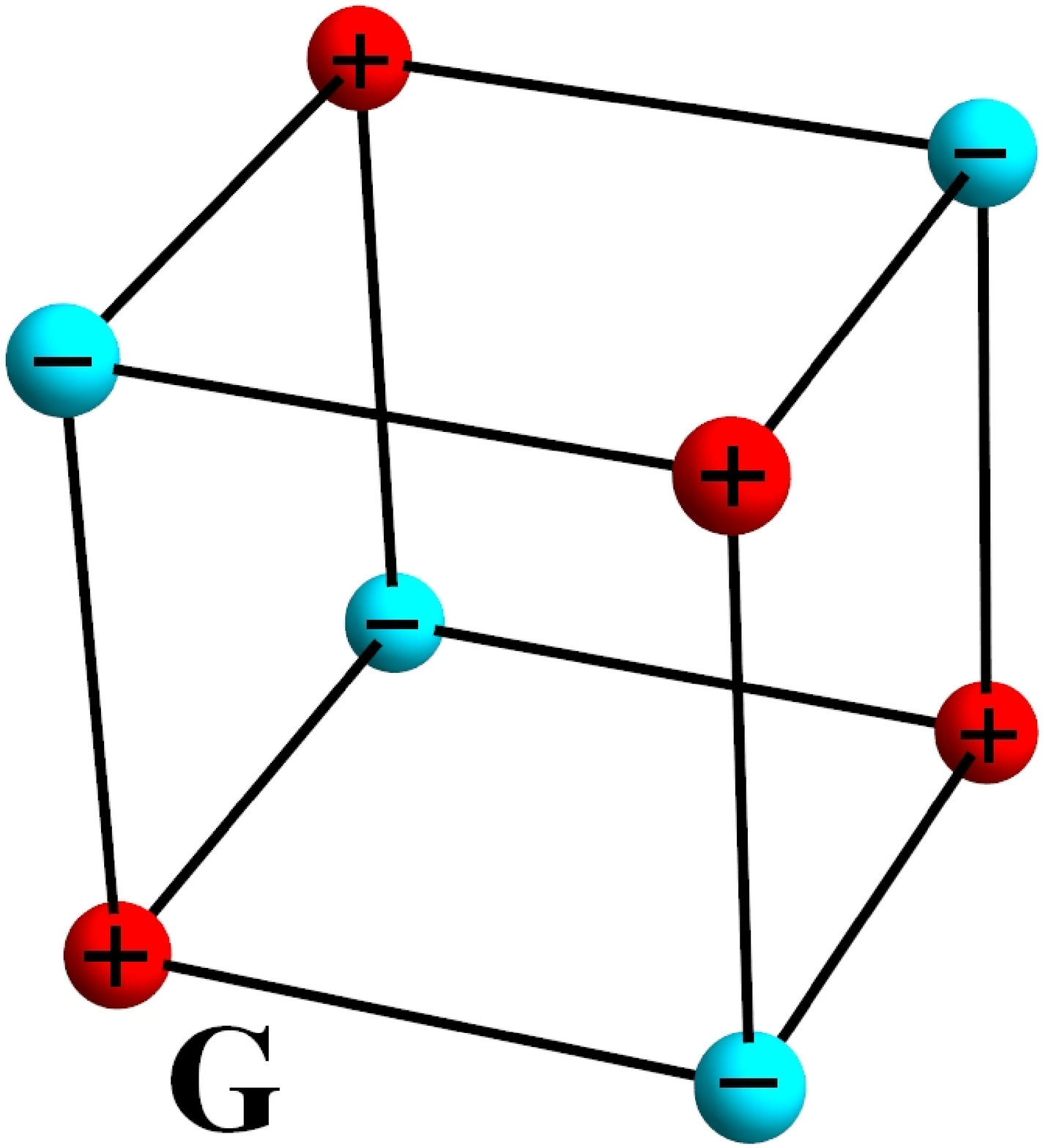}
\caption{(Color online) Types of magnetic ordering, as introduced by Wollan and Koehler\cite{Wollan}. Each sphere represents ${\rm Mn}$ ion and the sign/color represents the orientation of the z-axis spin projection. A-type describes planes with ferromagnetic alignment of spins, but with antiferromagnetic coupling between them. B-type is ferromagnetism, C-type corresponds to ferromagnetic alignment of spins along the chains, but antiferromagnetic in the plains, and G-type is the nearest-neighbor antiferromagnetism.}\label{fig2}
\end{figure*}

One of the most studied mixed-valence manganites is ${\rm La_{1-x}Ca_xMnO_3}$, which can be examined across the whole doping range $0 \leq x \leq 1$. While both of the end members ${\rm LaMnO_3}$ and ${\rm CaMnO_3}$ are antiferromagnetic and insulating, the resulting mixed-valence compound shows a variety of magnetic and transport properties depending on the value of the doping $x$. Wollan and Koehler \cite{Wollan} were the first to study the types of magnetic arrangements in the whole range of composition for ${\rm La_{1-x}Ca_xMnO_3}$ and to organize the results into a scheme of structures and structure transitions. At small hole densities, including $x = 0$, in addition to the already known pattern B, which corresponds to ferromagnetism, they observed pattern A. It corresponds to planes with ferromagnetic alignment of spins, but with antiferromagnetic coupling between those plains, and is known as A-type antiferromagnetism. Around $x \sim 0.8$, C-type pattern was observed, which corresponds to ferromagnetic alignment of spins along the chains, but antiferromagnetic in the plains. When $x$ approaches one, almost all of the manganese ions are in the $4+$ oxidation state and the dominant pattern is of G-type, with antiferromagnetism in all three directions. The corresponding types of magnetic arrangements are depicted in Figure \ref{fig2}.

As a typical member of the transition metals family, ${\rm Mn}$ has a incomplete $3d$ shell filled with four and three electrons for ${\rm Mn^{3+}}$ and ${\rm Mn^{4+}}$ respectively. Due to the crystal field splitting effect, the degeneracy of the five $3d$ orbitals is lifted and they are grouped into one triplet $t_{2g}$ and one doublet $e_g$. The triplet has lower energy because of the space orientation of the corresponding orbitals inside the octahedron of six oxygen ions, surrounding the central manganese ion. The population of the ${t}_{2g}$ electrons remains constant and
the Hund rule enforces alignment of the three $t_{2g}$ spins into a state of maximum spin $S=3/2$. Then, the ${t}_{2g}$ sector can be replaced by a \emph{localized spin} at each manganese ion, reducing the complexity of the original five orbital model. The electrons from the $e_g$ sector however can move from ion to ion, maintaining the projection of their spin, and are called \emph{mobile electrons}. The only important interaction between the two sectors is the Hund coupling between localized ${t}_{2g}$ spins and mobile ${e}_g$ electrons.

Zener \cite{Zener1,*Zener2} was the first to construct a model which correctly describes the properties of transition metals with incomplete $d$-shells, such as manganese. He introduced three principles, which he believed govern the interaction between the incomplete $d$-shells of neighboring atoms and later, as a demonstration of his ideas, he applied these principles to successfully  explain the connection between conductivity and ferromagnetism in manganites, observed by Jonker and Van Santen \cite{Jonker1,*Jonker2,*Jonker3}. He interpreted ferromagnetism as arising from the indirect coupling of incomplete $d$-shells via conducting electrons and also sketched a possible mechanism by which conduction electrons move between manganese ions. In this so-called ``double-exchange'' mechanism the transfer must occur through the oxygen ion between the two manganese ions, as a simultaneous ``double exchange'' of electrons. This process is a real charge transfer process and involves an overlap integral between the manganese $3d$ and oxygen's $2p$ orbitals. Because of the strong Hund's coupling, the transfer-matrix element has finite value only when the core spins of the ${\rm Mn}$ ions are aligned ferromagnetically. Zener also showed that ferromagnetism would never occur in the absence of conduction electrons or of some other indirect coupling.

The ideas behind the Double Exchange Model were further developed by Anderson and Hasegawa\cite{AH}, who showed that electron transfer between neighboring ${\rm Mn}$ ions depends on the angle between their magnetic moments as $t_{\rm eff} = t \cos(\theta/2)$. The transfer probability varies from one for $\theta =0$ to zero for $\theta = \pi$ and the exchange energy is lower when the itinerant electron's spin is parallel to the total spin of the {\rm Mn} cores. Starting from their result, de Gennes\cite{deGennes} introduced the so-called ``spin-canted'' state as a possible explanation of the coexistence of ferromagnetic and antiferromagnetic features. Another pioneer theoretical study was carried out by Goodenough\cite{Goodenough}, regarding the charge, orbital and spin arrangements in the non-ferromagnetic regime of the phase diagram of ${\rm La_{1-x}Ca_xMnO_3}$. It was based on the notions of ``semicovalent bond'' and elastic energy considerations.

Simple theories based on the Double Exchange Model however have several problems. They overestimate the Curie temperature of most manganites, cannot describe the huge magnitude of the Colossal Magnetoresistance effect, underestimate the resistivity values in the paramagnetic phase by several orders and cannot account for the existence of charge/orbital ordering, phase separation scenario and strong lattice effects/anomalies seen experimentally. Searching for a more elaborate models, Millis\cite{Millis1} was the first to argue that the physics of manganites is determined by the interplay between a strong electron-phonon interaction and the large Hund coupling. His ideas were later expanded by Roder\cite{Roder}, Zang\cite{Zang}, Yunoki\cite{Yunoki} and others.

The Jahn-Teller distortions lead to lifting the degeneracy of the $e_g$ orbitals, with energy difference of about 1 eV. The oxygens surrounding the manganese ion readjust their locations creating asymmetry between the different directions. This effectively removes the degeneracy of the $e_{g}$ orbitals. The lifting of the degeneracy due to the orbital-lattice interaction is called  Jahn-Teller effect. The splitting between the two $e_g$ orbitals is big enough to justify the use of the simple one-band Double Exchange Model for manganites as an acceptable approximation. More elaborate models should however consider the electron-phonon interactions explicitly and include the corresponding terms in the Hamiltonian of the system. The importance of the crystal structure is demonstrated by the result that, without the Jahn-Teller distortion (i.e. for a cubic cell), ${\rm LaMnO_3}$ would be a ferromagnetic metal rather than an A-type antiferromagnetic insulator\cite{Sarma,Pickett1,*Pickett2, Solovyev}. The Jahn-Teller effect responsible for the orthorhombic distortion also leads to the stabilization of the A-type antiferromagnetism.

The double exchange model with Jahn-Teller coupling is a widely used model for manganites. The procedures followed to obtain the essential features of the model are different: numerical studies \cite{Yunoki,Hotta}, Dynamical Mean-Field Theory (DMFT) \cite{Millis2,Millis3,Held}, \emph{ab initio} density-functional calculations \cite{Popovic}, and analytical calculations \cite{Millis2,*Millis3,Nolting1,*Nolting2}. In spite of the common conclusion that Jahn-Teller coupling suppresses the ferromagnetic state, the results are quite different and do not match the experimental results. For example the calculated Curie temperatures are two and even three times larger then the experimentally measured. Because of that it is important to formulate theoretical criteria for adequacy of the method of calculation. In our opinion the calculations should be in accordance with the Mermin-Wagner theorem \cite{M-W}. It claims that at nonzero temperature, a 1D or 2D isotropic spin-S Heisenberg model with finite-range exchange interaction can be neither ferromagnetic nor antiferromagnetic. We employ a technique of calculation \cite{Us2}, which captures the essentials of the magnon fluctuations in the theory, and for $2D$ systems one obtains zero Curie temperature, in accordance with Mermin-Wagner theorem. The physics of the ferromagnetic manganites near the Curie temperature is dominated by the magnon fluctuations and it is important to account for them in the best way. In contrast with other theories, our model includes the contribution of both the localized and the mobile electrons, thus better accounting for the magnon fluctuations. With our previous papers\cite{Us2,Us3} successfully explaining the qualitative picture, we now turn to comparing quantitative results and determining the model parameters which best describe the experimental data.

\section{Effective Model}

To construct the effective model, we start with the two-band Double Exchange Model with added Jahn-Teller distortion and antiferromagnetic (Heisenberg) term. The Hamiltonian reads \be H = H_{\rm DE}  + H_{\rm JT} + H_{\rm AF}.\label{H}\ee The first term describes the hopping of $e_g$ electrons and the Hund interaction between the spin $\textbf{s}_i$ of the $e_g$ electron and the localized $t_{2g}$ spin $\textbf{S}_i$ \be\label{JT2}H_{\rm DE} = -\sum\limits_{i\, \textbf{a}\,ll'\alpha } {t_{ll'}^{\textbf{a}} c_{i l\alpha }^ +  c_{i+\textbf{a}\, l'\alpha } }- 2J_H \sum\limits_i {\textbf{s}_i \cdot \textbf{S}_i } \ee where $c_{i l\alpha }^+ $ and $c_{i l\alpha }$ are creation and annihilation operators for $e_g$ electron with spin $\alpha$ on orbitals $d_{x^2-y^2}(l=a)$ and $d_{3z-r^2}(l=b)$ at site $i$. The sums are over all sites of a three-dimensional cubic lattice, and $\textbf{a}$ is the vector connecting nearest-neighbor sites. For the cubic lattice, the hopping amplitudes between the orbitals along the $x,y,z$ directions are: \bea
t_{aa}^x  &=&  - \sqrt 3 t_{ab}^x  =  - \sqrt 3 t_{ba}^x  = 3t_{bb}^x  = t\nonumber\\
t_{aa}^y  &=& \sqrt 3 t_{ab}^y  = \sqrt 3 t_{ba}^y  = 3t_{bb}^y  = t\\
t_{aa}^z  &=& t_{ab}^z  = t_{ba}^z  = 0,\quad t_{bb}^z  = 4t/3\nonumber
\eea

The second term in \eqref{JT2} is the Hund interaction between the spin $\textbf{s}_i$ of the $e_g$ electron and the localized $t_{2g}$ spin $\textbf{S}_i$ with
\be s^{\nu}_i = \frac 12\sum\limits _{l\alpha\beta} c_{i l\alpha }^ +\sigma^{\nu}_{\alpha\beta} c_{il\beta},\ee where $\sigma^x,\sigma^y,\sigma^z$ are the Pauli matrices, and the Hund's constant $J_{\rm H}$ is positive.

The $H_{\rm JT}$ part models the coupling of the $e_g$ electrons to the phonons: \be\label{JT3} H_{\rm JT}  = g\!\sum\limits_i \left(Q_{2i} \tau _{xi}\! + Q_{3i} \tau _{zi} \right)+ \frac{k}{2}\sum\limits_i {\left( Q_{2i}^2  + Q_{3i}^2 \right)}\ee where \bea\tau _{xi}  = \sum\limits_{\alpha}  \left( c_{ia\alpha }^ +  c_{ib\alpha }  + c_{ib\alpha }^ +  c_{ia\alpha } \right)\nonumber\\
\tau _{zi}  = \sum\limits_{\alpha} \left(c_{ia\alpha }^ +  c_{ia\alpha}  - c_{ib\alpha}^ +  c_{ib\alpha}\right)\eea are the so-called pseudo-spin operators, $g$ is the electron-phonon coupling constant, and $Q_{2i}$ and $Q_{3i}$ are the Jahn-Teller phonon modes. The second term in $H_{\rm JT}$ is the general quadratic potential for distortions with constant $k$. The important energy scale of the phonon-electron interaction is the static Jahn-Teller energy $E_{\rm JT}=g^2/(2k)$.

To explain the experimentally observed antiferromagnetic phases in manganites we will also need a Heisenberg-like antiferromagnetic term $H_{\rm AF}$, which can be represented as: \be{H_{\rm AF}} = \sum\limits_{ ia } {J_{\rm AF}^{\bf a}}{\bf S}_i \cdot {\bf S}_{i+{\bf a}},\label{AF}\ee where we have different exchange constants $J_{\rm AF}$ for different lattice directions.

We switch to Schwinger-boson representation for the localized spin operators $\textbf{S}_i$ \be S^{\nu}_i  = \frac{1}{2}\sum\limits_{\alpha \beta}\varphi _{i\alpha }^ +  \sigma^{\nu}_{\alpha \beta} \varphi_{i\beta},\qquad {\textrm{ with }}\sum\limits_{\alpha}\varphi _{i\alpha }^ + \varphi _{i\alpha}\!=\!2s.\ee

By means of the Schwinger-bosons we introduce spin-singlet Fermi fields
\bea\label{Fsinglet} &&\hskip -1cm\Psi^A_{il}(\tau)=\frac {1}{\sqrt {2s}}\left[\varphi^+_{i 1}(\tau)c_{il1}(\tau)+\varphi^+_{i 2}(\tau)c_{il2}(\tau)\right]\label{cfm8}\\
&& \hskip -1cm \Psi^B_{il}(\tau)=\frac {1}{\sqrt{2s}}\left[\varphi_{i1}(\tau)c_{il2}(\tau)\,-\,\varphi_{i2}(\tau)c_{il1}(\tau)\right] \label{cfm9} \eea
and write the spin of the $e_g$ electron and the total spin of the system \be\textbf{S}^{\rm tot}_{i} =  \textbf{S}_i+\textbf{s}_{i}\ee
in terms of the singlet fermions \be {\bf S}^{\rm tot}_{i}  = \frac 1s  \left[s+\frac 12 \sum\limits_{il}\left (\Psi^{+A}_{il}\Psi^A_{il}-\Psi^{+B}_{il}\Psi^B_{il}\right)\right]{\bf S}_i, \ee where the above two formulas account for the fact that the spins of the $e_g$ and $t_{2g}$ electrons are parallel.

If we average  the total spin of the system in the subspace of the singlet fermions $A$ and $B$, the vector \be \textbf{M}_i= \langle \textbf{S}^{\rm tot}_i \rangle _f\ee identifies
the local orientation of the total magnetization. Because of the fact that $t_{2g}$-electron spin is parallel with $e_{g}$-electron spin we obtain
$\textbf{M}_i = \frac MS \textbf{S}_i$ with \be M = S+\frac 12 \sum\limits_{il} \langle \left (\Psi^{A+}_{il}\Psi^A_{il}-\Psi^{B+}_{il}\Psi^B_{il}\right) \rangle_f.\ee If we use Holstein-Primakoff representation for the vectors $\textbf{M}_i$ with $M$ as an ``effective spin'' of the system $(\textbf{M}_i^2=M^2)$, \bea\label{e:3.20} &&M_j^+ = M_{j1} + i M_{j2}=\sqrt {2M-a^+_ja_j}\,\,\,\,a_j \nonumber \\
&&M_j^- = M_{j1} - i M_{j2}=a^+_j\,\sqrt {2M-a^+_ja_j} \\[4pt] &&M^{3\;}_j =2M - a^+_j a_j, \nonumber \eea the Bose fields $a_i$ and $a^+_i$ are the \textbf{true magnons} of the system.

An important advantage of working with singlet fermions is the fact that in terms of these spin-singlet fields the spin-fermion interaction is in a diagonal form, the spin variables (magnons) are
removed, and one accounts for it exactly: \be \sum\limits_{il} s_{il} \cdot S_{i}  = \frac{s}{2}\sum\limits_{il} \left( {\Psi
_{il}^{ + A} \Psi _{il}^A - \Psi _{il}^{ + B} \Psi _{il}^B } \right). \ee

Invoking (\ref{cfm8}-\ref{cfm9}) we can also rewrite the JT term as \bea H_{\rm JT} = g\sum\limits_i \Big[ Q_{2i} \left( \Psi _{i\sigma a}^ +  \Psi _{i\sigma b}  + \Psi _{i\sigma b}^ +  \Psi _{i\sigma a}  \right) \\
\hskip -.5cm +\,Q_{3i} \left( {\Psi _{i\sigma a}^ +  \Psi _{i\sigma a}-\Psi _{i\sigma b}^ +  \Psi _{i\sigma b} } \right)\!\Big]+\frac{k}{2}\sum\limits_i {\left( Q_{2i}^2  + Q_{3i}^2 \right)}\nonumber \eea

The theory is quadratic with respect to the spin-singlet fermions and one can integrate them out to obtain the free energy of fermions as a function of the magnons' fields. We expand the free energy in the ferromagnetic regime in powers of magnons' fields and keep only the first two terms. The first term ${\cal F}_{f0}$, which does not depend on the magnons' fields, is the free energy of Fermions with spins of localized $t_{2g}$ electrons treated classically. We fix the model parameters and consider this term as a function of the Jahn-Teller distortion modes independent on the lattice sites. One can then show that this term depends only on $\sqrt{Q_2^2+Q_3^2}$. If we represent $Q_2=\hat Q \cos\gamma$ and $Q_3=\hat Q \sin\gamma $, this allows us to fix $\gamma=0$ $(Q_{3}=0)$. The physical value of the Jahn-Teller distortion is the value at which ${\cal F}_{f0}$ has a minimum. In this way we obtain the distortion as a function of the density of $e_g$ electrons for different values of the Jahn-Teller energy and fixed Hund's coupling\cite{Us3}.

The Hamiltonian corresponding to the free fermion part has the form \bea H_{\rm f} = \sum\limits_k \Big[\varepsilon _{ak}^A\Psi _{ak}^{ + A}\Psi _{ak}^A + \varepsilon _{ak}^B\Psi _{ak}^{ + B}\Psi _{ak}^B \nonumber\\
+\varepsilon _{bk}^A\Psi _{bk}^{ + A}\Psi _{bk}^A +\varepsilon _{bk}^B\Psi _{bk}^{ + B}\Psi _{bk}^B - \varepsilon (k)\Psi _{ak}^{ + A}\Psi _{bk}^A \nonumber\\[8pt]
-\varepsilon (k)\Psi _{ak}^{ + B}\Psi _{bk}^B - \varepsilon (k)\Psi _{bk}^{ + A}\Psi _{ak}^A - \varepsilon (k)\Psi _{bk}^{ + B}\Psi _{ak}^B \Big],
\eea where \bea
&  \varepsilon _{ak}^A  =  - 2t\left( {\cos k_x  + \cos k_y } \right) - sJ_{\rm H} - \mu\nonumber    \\[10pt]
&  \varepsilon _{ak}^B  =  - 2t\left( {\cos k_x  + \cos k_y } \right) + sJ_{\rm H} - \mu   \nonumber \\[10pt]
&\displaystyle  \varepsilon _{bk}^A  =  - 2t\left( {\frac{1}{3}\cos k_x  + \frac{1}{3}\cos k_y  + \frac{4}{3}\cos k_z } \right) - sJ_{\rm H} - \mu  \nonumber\\[8pt]
&\displaystyle \varepsilon _{bk}^B  =  - 2t\left( {\frac{1}{3}\cos k_x  + \frac{1}{3}\cos k_y  + \frac{4}{3}\cos k_z } \right) + sJ_{\rm H} - \mu   \nonumber\\[5pt]
&\displaystyle  \varepsilon (k) = 2\left( { - \frac{t}{{\sqrt 3 }}\cos k_x  + \frac{t}{{\sqrt 3 }}\cos k_y  - \frac{{g Q}}{2}} \right)\eea
are the dispersions of the spin-singlet fermions. In order to diagonalize this Hamiltonian we use a Bogolyubov-like transformation

\be \left| \begin{array}{l}
   {\Psi _{ak}^A  = u_k^A f_{ak}^A  + v_k^A f_{bk}^A }  \\[10pt]
   {\Psi _{bk}^A  =  - v_k^A f_{ak}^A  + u_k^A f_{bk}^A }\\[10pt]
   {\Psi _{ak}^B  = u_k^B f_{ak}^B  + v_k^B f_{bk}^B }  \\[10pt]
   {\Psi _{bk}^B  =  - v_k^B f_{ak}^B  + u_k^B f_{bk}^B }
\end{array} \right.\ee with coefficients \be\begin{array}{l}
\displaystyle u^R_k  = \sqrt {(1 + x^R_k)/2} \qquad \qquad R=A,B\\[10pt]
\displaystyle v^R_k = {\mathop{ {\rm sign} (\varepsilon(k))}} \sqrt {(1 - x^R_k)/2}\\[10pt]
\displaystyle x^R_k  = \frac{\varepsilon _{bk}^R - \varepsilon _{ak}^R }{\sqrt{4\varepsilon ^2 (k) + \left( {\varepsilon _{bk}^R - \varepsilon _{ak}^R } \right)^2 }}\end{array}\ee

The Hamiltonian can then be rewritten in diagonal form in terms of the quasiparticles

\bea H_{\rm f}  = \sum\limits_k \Big[ E_{ak}^A f_{ak}^{ + A}f_{ak}^A + E_{ak}^B f_{ak}^{ + B} f_{ak}^B  \nonumber\\
+ E_{bk}^A f_{bk}^{ + A} f_{bk}^A  + E_{bk}^B f_{bk}^{ + B} f_{bk}^B  \Big] \eea

where the dispersions of the newly introduced fermions $f$, $f^+$ are given by \bea
&\displaystyle  E_{ak}^A  = \frac{{\varepsilon _{bk}^A  + \varepsilon _{ak}^A }}{2} - \frac{1}{2}\sqrt {4\varepsilon ^2 (k) + \left( {\varepsilon _{bk}^A  - \varepsilon _{ak}^A } \right)^2 }  \nonumber \\[10pt]
&\displaystyle  E_{bk}^A  = \frac{{\varepsilon _{bk}^A  + \varepsilon _{ak}^A }}{2} + \frac{1}{2}\sqrt {4\varepsilon ^2 (k) + \left( {\varepsilon _{bk}^A  - \varepsilon _{ak}^A } \right)^2 }  \nonumber\\[10pt]
&\displaystyle  E_{ak}^B  = \frac{{\varepsilon _{bk}^B  + \varepsilon _{ak}^B }}{2} - \frac{1}{2}\sqrt {4\varepsilon ^2 (k) + \left( {\varepsilon _{bk}^B  - \varepsilon _{ak}^B } \right)^2 }  \nonumber\\[10pt]
&\displaystyle  E_{bk}^B  = \frac{{\varepsilon _{bk}^B  + \varepsilon _{ak}^B }}{2} + \frac{1}{2}\sqrt {4\varepsilon ^2 (k) + \left( {\varepsilon _{bk}^B  - \varepsilon _{ak}^B } \right)^2 }
\eea

The second term in the Fermion free energy gives the spin-fermion interaction

\begin{widetext}\be\begin{array}{r}\displaystyle H_{\rm s - f} = -\sum\limits_{< ij> }\frac{t_{aa}^{<ij>}}{2s} \bigg[\left( {\varphi_{i\sigma}^+ \varphi_{j\sigma}-2s} \right)\left( {\Psi _{ai}^{ + A} \Psi _{aj}^A  + \Psi _{aj}^{ + B} \Psi _{ai}^B } \right) + \left({\varphi_{j\sigma}^+ \varphi_{i\sigma}-2s} \right)\left( {\Psi _{ai}^{+ B}\Psi _{aj}^B  + \Psi _{aj}^{ + A} \Psi _{ai}^A } \right) \\
+\displaystyle \left( {\varphi _{ai}^ + \varphi _{bj}^ +   - \varphi _{aj}^ +  \varphi _{bi}^ +  } \right)\left( {\Psi _{aj}^{+ A} \Psi _{ai}^B  - \Psi _{ai}^{ + A} \Psi _{aj}^B } \right) +
\left( {\varphi _{ai} \varphi _{bj}  - \varphi _{bi}\varphi _{aj}} \right)\left( {\Psi _{ai}^{ + B} \Psi_{aj} ^A  - \Psi _{aj}^{ + B} \Psi _{ai}^A } \right)\bigg] \\
-\displaystyle\sum\limits_{< ij> } \frac{t_{bb}^{<ij>}}{2s} \bigg[\left({\varphi_{i\sigma}^+ \varphi_{j\sigma}-2s} \right)\left( {\Psi _{bi}^{ + A} \Psi _{bj}^A  + \Psi _{bj}^{ + B} \Psi _{bi}^B }\right) + \left( {\varphi_{j\sigma}^+ \varphi_{i\sigma}-2s} \right)\left( {\Psi _{bi}^{+ B} \Psi _{bj}^B  + \Psi _{bj}^{ + A} \Psi _{bi}^A } \right) \\
+\displaystyle  \left( {\varphi _{ai}^ + \varphi _{bj}^ +   -\varphi _{aj}^ +  \varphi _{bi}^ +  } \right)\left( {\Psi _{bj}^{ + A} \Psi _{bi}^B  - \Psi _{bi}^{ + A} \Psi _{bj}^B } \right) +
\left({\varphi _{ai} \varphi _{bj}  - \varphi _{bi} \varphi _{aj}} \right)\left( {\Psi _{bi}^{ + B} \Psi_{bj}^A  - \Psi _{bj}^{ + B} \Psi _{bi}^A } \right)\bigg] \\
-\displaystyle \sum\limits_{< ij> }\frac{t_{ab}^{<ij>}}{2s}  \bigg[ \left( {\varphi_{i\sigma}^+ \varphi _{j\sigma} - 2s} \right)\left( {\Psi _{ai}^{ + A} \Psi _{bj}^A+\Psi _{bi}^{ + A} \Psi _{aj}^A  +\Psi _{bj}^{ + B} \Psi _{ai}^B+\Psi _{aj}^{ + B} \Psi_{bi}^B}\right)\\[3pt]
+\displaystyle \left( {\varphi _{j\sigma }^+ \varphi_{i\sigma}-2s} \right) \left( {\Psi _{ai}^{+ B} \Psi _{bj}^B+\Psi _{bi}^{+B} \Psi _{aj}^B +\Psi _{bj}^{ + A} \Psi _{ai}^A + \Psi _{aj}^{ + A} \Psi _{bi}^A } \right)\\[8pt]
+\displaystyle  \left( {\varphi _{ai}^ + \varphi _{bj}^ +   -\varphi _{aj}^ +  \varphi _{bi}^ +  } \right)\left( {\Psi _{aj}^{+ A} \Psi _{bi}^B+\Psi _{bj}^{ + A} \Psi _{ai}^B -\Psi _{ai}^{ +
A} \Psi _{bj}^B - \Psi _{bi}^{ + A} \Psi _{aj}^B } \right)\\[3pt]
+\displaystyle \left( {\varphi _{ai} \varphi _{bj}  - \varphi_{bi} \varphi _{aj} }\right)\left( {\Psi _{bi}^{ + B} \Psi _{aj}^A+ \Psi_{ai}^{ + B} \Psi _{bj}^A - \Psi _{aj}^{ + B} \Psi _{bi}^A
-\Psi _{bj}^{ + B} \Psi _{ai}^A} \right)\bigg] \end{array}\ee\end{widetext}

The spin-fermion Hamiltonian is quadratic with respect to the magnons' fields and defines the effective magnon Hamiltonian in Gaussian approximation:

\be\label{JTeff} h_{\rm eff}^{it} = \sum\limits_{i \textbf{a}} \rho^{\textbf{a}}\left(a_i^+a_i + a_{i+\textbf{a}}^+a_{i+\textbf{a}} - a_i^+a_{i+\textbf{a}} - a_{i+\textbf{a}}^+a_i \right)\ee

Based on the rotational symmetry, one can supplement this Hamiltonian up to an effective Heisenberg-like Hamiltonian, written in terms of the vectors $\textbf{M}_i$

\be H_{\rm eff} = -\sum\limits_{ia}J^{\bf a} {\bf M}_i \cdot {\bf M}_{i+{\bf a}},\label{JTeff2}\ee
where $J^{{a}}$ are the effective couplings \be J^a = -\frac{S^2}{M^2}J_{\rm AF}^a + \frac{\rho^a}{M},\ee

The effective exchange constants depend on the lattice direction and are a sum of two terms. The first gives the contribution of the antiferromagnetic Hamiltonian \eqref{AF}, rewritten in terms of the vectors ${\bf M}_i$. The second term gives the contribution from the spin-fermion interaction, where $\rho^{\textbf{a}}$ are the spin-stiffness constants \eqref{rho}, which also depend on the space directions $\textbf{a}$. They are calculated at zero temperature for fixed Hund's coupling, Jahn-Teller energy, charge density, and Jahn-Teller distortion determined above using the technique introduced in previous papers\cite{Us2}, and have the form

\begin{widetext}\bea \rho^{{\mu}}  =  \frac{1}{V} \sum\limits_k\Bigg[\frac{t_{ab}^{{\mu}}}{M}  \Big( u_k^A v_k^A \left( n_{bk}^A - n_{ak}^A  \right) + u_k^B v_k^B \left( n_{bk}^B   -n_{ak}^B \right)\Big) +\nonumber\\
 +\frac{t_{aa}^{\mu}}{2M}\Big(  \left(u_k^A\right)^2 n_{ak}^A + \left( v_k^A\right)^2 n_{bk}^A  + \left(u_k^B\right)^2 n_{ak}^B + \left(v_k^B\right)^2 n_{bk}^B \Big) +\nonumber\\
 +\frac{t_{bb}^{{\mu}}}{2M}\Big(  \left(u_k^A\right)^2 n_{bk}^A + \left( v_k^A\right)^2 n_{ak}^A  + \left(u_k^B\right)^2 n_{bk}^B + \left(v_k^B\right)^2 n_{ak}^B \Big) \Bigg]\cos k_{{\mu}} +\nonumber\\
+ \frac{2}{M} \frac{1}{V} \sum\limits_k \Bigg[\left( {t_{aa}^{{\mu}} u_k^A u_k^B + t_{bb}^{{\mu}} v_k^A v_k^B  - t_{ab}^{{\mu}}u_k^A v_k^B  - t_{ab}^{{\mu}}v_k^A u_k^B } \right)^2  \left(\frac{{n_{ak}^B - n_{ak}^A}}{{E_{ak}^B  - E_{ak}^A }}\right)  +\nonumber\\
 + \left( {t_{aa}^{{\mu}} u_k^A v_k^B  - t_{bb}^{{\mu}} v_k^A u_k^B  +t_{ab}^{{\mu}}u_k^A u_k^B  - t_{ab}^{{\mu}}v_k^A v_k^B } \right)^2  \left(\frac{{n_{bk}^B - n_{ak}^A}}{{E_{bk}^B  - E_{ak}^A }}\right) +  \nonumber\\
+ \left( {t_{aa}^{{\mu}} v_k^A u_k^B  - t_{bb}^{{\mu}} u_k^A v_k^B -t_{ab}^{{\mu}}v_k^A v_k^B  + t_{ab}^{{\mu}}u_k^A u_k^B } \right)^2 \left(\frac{{n_{ak}^B  - n_{bk}^A}}{{E_{ak}^B  - E_{bk}^A }}\right) +\nonumber\\
+ \left( {t_{aa}^{{\mu}} v_k^A v_k^B  + t_{bb}^{{\mu}} u_k^A u_k^B  + t_{ab}^{{\mu}}v_k^A u_k^B  + t_{ab}^{{\mu}}u_k^A v_k^B } \right)^2 \left(\frac{{n_{bk}^B - n_{bk}^A }}{{E_{bk}^B  - E_{bk}^A }}\right) \Bigg]\sin^2 k_{{\mu}} \label{rho}\eea
\end{widetext} where $n^R_{lk}$ are the occupation numbers for the $f_{lk}^R$ quasiparticles respectively.

The ferromagnetic phase is stable if all effective exchange coupling constants are positive $J^{{a}} > 0$. If one of them is negative, for example $J^y<0$, and the other two are positive $J^x>0$, $J^z>0$, the stable state is A-type antiferromagnetic phase which has planes $(x,z)$ that are ferromagnetic (parallel moments), with antiferromagnetic (antiparallel) arrangement of the magnetic moments between them (see fig.\;\ref{fig2}).

\begin{figure*}[!ht]
\includegraphics[width=\linewidth]{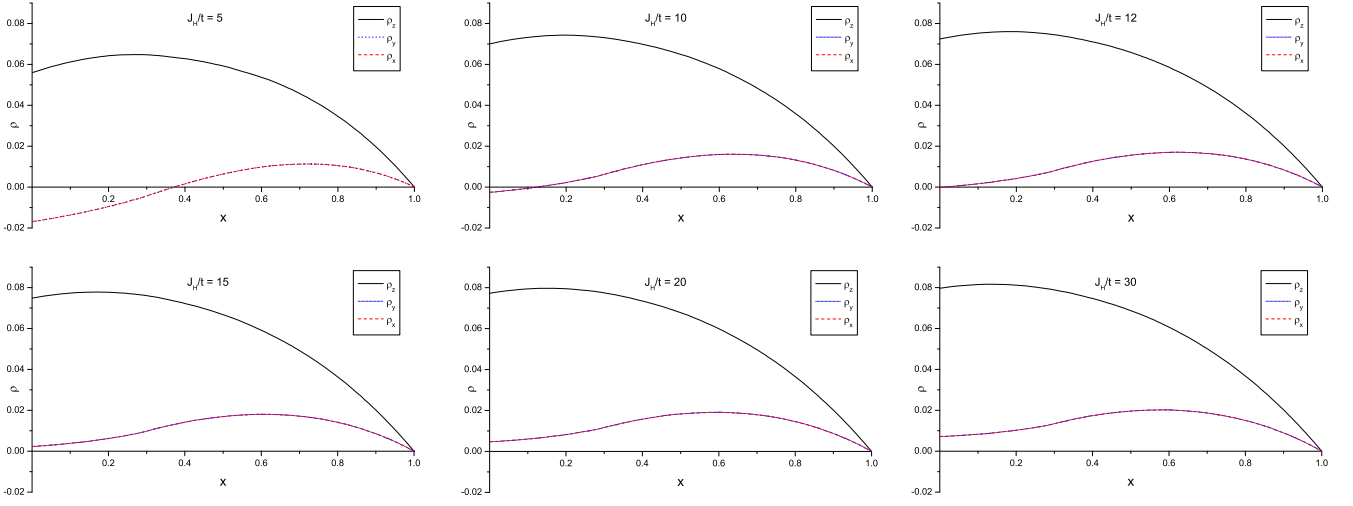}
\caption{(Color online) Spin-stiffness constants $\rho_a$ as a function of hole doping $x$ for different values of Hund's coupling $J_{\rm H}/t$ in pure Double Exchange Model ($J_{\rm AF}^a = 0$, $E_{\rm JT} = 0$). In this simple case $\rho_x$ and $\rho_y$ are equal across all doping values.}\label{fig4}
\end{figure*}

The spin-stiffness constants, as a function of the doping $x$, are depicted in Figure \ref{fig4} for the pure Double Exchange Model ($J_{\rm AF}^a = 0$, $E_{\rm JT} = 0$), across a broad range of values for the Hund's coupling $J_{\rm H}/t$. In such simple model we account only for the contribution of itinerant electrons, and as a consequence $\rho_x$ and $\rho_y$ remain equal for all carrier densities. One can see that by increasing the value of $J_{\rm H}/t$ the spin-stiffness constants also increase, with $\rho_z$ being the dominant one. Consequently, the range of the ferromagnetic region, determined in this case by the condition $\rho^a >0$ for all $a$, also increases with increase of $J_{\rm H}/t$. From the form of the spin-stiffness curves we can estimate a limit for the value of the Hund's coupling, needed to reproduce the observed phases. Since the ferromagnetic to A-type antiferromagnetic transition occurs around $x\!=\!0.08$, it is clear that values of $J_{\rm H}/t < 10$ are not acceptable. The minimum value for $J_{\rm H}/t$ that extends the ferromagnetic phase to and beyond $x=0.08$ is around $J_{\rm H}/t=12$, in accordance with previous results\cite{Us1,Golosov,Pekker}. For a realistic models, values as big as $J_{\rm H}/t=20$ should be used to ensure that we will be well above the limit, since the addition of both the antiferromagnetic and Jahn-Teller terms leads to suppression of the magnon fluctuations\cite{Us1,Us3}. Another important observation we can make about Figure \ref{fig4} is that the maximum of the spin-stiffness curves is shifted to smaller values of $x$ with increase of $J_{\rm H}/t$. This in turn determines the behavior of the Curie temperature curves and will play important role in determining the model parameters later. We will come back to this observation once we have introduced the method of calculating the Curie temperatures.

The simple model used to construct Figure \ref{fig4} however cannot explain all the observed phases. For example, it is a well known fact that the $\rm CaMnO_3$ end member of the ${\rm La_{1-x}Ca_xMnO_3}$ family is a G-type antiferromagnet (see Fig.\;\ref{fig2}). In addition to the G-type antiferromagnetic phase, with decreasing the doping C-type antiferromagnetic arrangement is observed before reaching the more complicated CE phase\cite{Cheong,Fuji1,Fuji2}. Both the G-type and the C-type antiferromagnetic arrangements require the Double Exchange Model Hamiltonian \eqref{JT2} to be supplemented with antiferromagnetic term \eqref{AF}. However, from Figure 3 it is evident that antiferromagnetism cannot be added in a trivial way. If we have equal value of $J_{\rm AF}$ for all three lattice directions, to have G-type antiferromagnetism this value has to be larger than $\rho_z/M$. As a result the other two spin-stiffness constants will be negative for all doping values. Thus, if we want to describe the correct sequence of phase transitions $A\to B \to CE\to C\to G$ phase, we need to have different values of $J_{\rm AF}$ in different lattice directions, with $J_{\rm AF}^z$ being the largest. To explain the existence of the A-type antiferromanetic phase, having in mind that $\rho_x=\rho_y$ for all doping values, one might also consider having different values for $J_{\rm AF}^x$ and $J_{\rm AF}^y$. However, the addition of Jahn-Teller distortions splits $\rho_x$ and $\rho_y$, as discussed below, and we can describe the existence of the A-type antiferromagnetic phase even when $J_{\rm AF}^x$=$J_{\rm AF}^y$. For simplicity we choose to work with equal values for them, which in turn leads to equal values of the effective couplings $J^x$ and $J^y$ across wide range of doping values, up to the appearance of Jahn-Teller effects. The importance of different values of $J_{\rm AF}$ in different lattice directions will become more evident when we discuss the phase portraits.

\begin{figure}[!b]
\includegraphics[width=.98\linewidth]{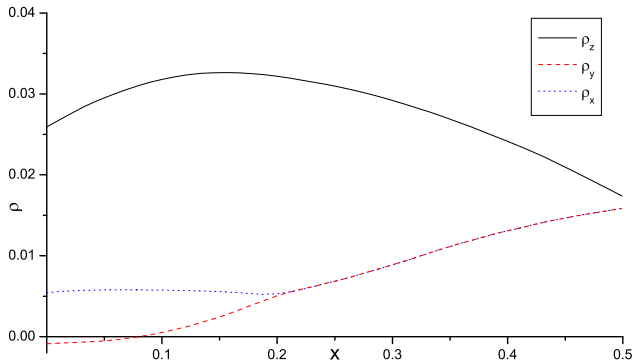}
\caption{(Color online) Spin-stiffness constants $\rho_a$ as a function of hole doping $x$ in the presence of Jahn-Teller distortion.}\label{fig5}
\end{figure}

To help illustrate the impact of the Jahn-Teller distortions on the spin stiffness constants, we have depicted them in Figure \ref{fig5} for the case of non-zero $E_{\rm JT}$. In accordance with the experimental\cite{Kawano} and theoretical\cite{Us3} results, we have chosen a value for the Jahn-Teller energy with diminishing contribution beyond $x = 0.1$. One can see that the appearance of distortions is accompanied with a change of the slopes of the spin-stiffness curves. The distortion splits the $\rho^y$ (dashed) and $\rho^x$ (dotted) lines, causing $\rho^y$ to decrease rapidly, while $\rho_x$ is stabilized. At the critical density $x=0.08$, $\rho_y$ becomes equal to zero and the system undergoes a transition from ferromagnetic phase to A-type antiferromagnetic phase. Thus our model requires the existence of Jahn-Teller distortions to correctly describe the transition from ferromagnetic to A-type antiferromagnetic phase. The other remaining spin-stiffness constant, $\rho_z$, also starts to decrease. Since the spin stiffness constants are a measure for the magnon fluctuations in the ferromagnetic phase, which in turn determine the Curie temperature, the addition of Jahn-Teller distortions leads to decreasing the Curie temperature in the ferromagnetic regime\cite{Us3}.

\section{Critical temperature in the ferromagnetic regime}

To calculate the Curie temperature $T_C$ we utilize Schwinger-bosons mean-field theory \cite{S-b1}. The advantage of this method of calculation is that for 2D systems one obtains zero Curie temperature, in accordance with the Mermin-Wagner theorem\cite{M-W}. We start by representing the vector $\textbf{M}_i$ by means of Schwinger bosons ($\phi_{i\alpha},\phi_{i\alpha}^+$)

\be M^{\nu}_i = \frac{1}{2}\sum\limits_{\alpha \beta} {\phi _{i\alpha }^+ \sigma^{\nu}_{\alpha \beta} } \phi _{i\beta}\qquad\phi
_{i\alpha}^+ \phi _{i\alpha} =2M\ee
Next we use the identity \be \textbf{M}_i \cdot \textbf{M}_j  = \frac{1}{2}\left( {\phi _{i\alpha}^ +  \phi_{j\alpha} } \right)\left( {\phi _{j\beta}^ + \phi _{i\beta} } \right)-\frac{1}{4}\left( {\phi _{i\alpha}^ +  \phi _{i\alpha} }\right)\left( {\phi _{j\beta}^ +  \phi _{j\beta} }\right)\label{eqM}\ee
and rewrite the effective Hamiltonian in the form
\be\label{JTeff3} H_{\rm eff}  =  - \frac{1}{2}\sum\limits_{i\textbf{a} }J^{\textbf{a}} {\left( {\phi _{i\alpha
}^ +  \phi _{i+\textbf{a}\alpha} } \right)\left( {\phi _{i+\textbf{a}\beta}^ +  \phi
_{i\beta} } \right)}\ee where the constant term is dropped. To ensure the Schwinger boson constraint we introduce a parameter ($\lambda$) and add a new term to the effective Hamiltonian \eqref{JTeff3}. \be \label{JTeff4} \hat H _{\rm eff} = H_{\rm eff}  + \lambda \sum\limits_i {\left( {\phi _{i\sigma }^ + \phi _{i\sigma }  - 2M} \right)}\ee

We treat the four-boson interaction within Hartree-Fock approximation. The Hartree-Fock hamiltonian which corresponds to the effective hamiltonian reads

\begin{widetext}
\bea {H_{\rm HF}} = \frac{J}{2}\sum\limits_i \Big[ {J_x}\left\langle {\varphi _{i\sigma }^ + {\varphi _{i + {e_x}\sigma }}} \right\rangle \left\langle {\varphi _{i + {e_x}\sigma'  - }^ + {\varphi _{i\sigma '}}} \right\rangle + {J_y}\left\langle {\varphi _{i\sigma }^ + {\varphi _{i + {e_y}\sigma }}} \right\rangle \left\langle {\varphi _{i + {e_y}\sigma '}^ + {\varphi _{i\sigma '}}} \right\rangle + {J_z}\left\langle {\varphi _{i\sigma }^ + {\varphi _{i + {e_z}\sigma }}} \right\rangle \left\langle {\varphi _{i + {e_z}\sigma '}^ + {\varphi _{i\sigma '}}} \right\rangle\nonumber\\
- {J_x}\left\langle {\varphi _{i\sigma '}^ + {\varphi _{i + {e_x}\sigma '}}} \right\rangle \varphi _{i + {e_x}\sigma }^ + {\varphi _{i\sigma }} - {J_x}\left\langle {\varphi _{i + {e_x}\sigma '}^ + {\varphi _{i\sigma '}}} \right\rangle \varphi _{i\sigma }^ + {\varphi _{i + {e_x}\sigma }} - {J_y}\left\langle {\varphi _{i\sigma '}^ + {\varphi _{i + {e_y}\sigma '}}} \right\rangle \varphi _{i + {e_y}\sigma }^ + {\varphi _{i\sigma }}\nonumber\\
- {J_y}\left\langle {\varphi _{i + {e_y}\sigma '}^ + {\varphi _{i\sigma '}}} \right\rangle \varphi _{i\sigma }^ + {\varphi _{i + {e_y}\sigma }} - {J_z}\left\langle {\varphi _{i\sigma '}^ + {\varphi _{i + {e_z}\sigma '}}} \right\rangle \varphi _{i + {e_z}\sigma }^ + {\varphi _{i\sigma }} - {J_z}\left\langle {\varphi _{i + {e_z}\sigma '}^ + {\varphi _{i\sigma '}}} \right\rangle \varphi _{i\sigma }^ + {\varphi _{i + {e_z}\sigma }}   + \lambda \sum\limits_i {\left( {\varphi _{i\sigma }^ + {\varphi _{i\sigma }} - 2M} \right)}\quad \eea\end{widetext}

It can be rewritten in more compact form as \bea \label{JTH-F} & & H_{\rm H - F}= \frac{1}{2}\sum\limits_{i\textbf{a}} J^{\textbf{a}}\bar u_{i,i+\textbf{a}} u_{i,i+\textbf{a}} +\lambda \sum\limits_i {\left( {\phi _{i\sigma }^ +  \phi _{i\sigma }  - 2M} \right)}\nonumber \\
& - & \frac{1}{2}\sum\limits_{i\textbf{a}} J^{\textbf{a}}{\left[ {\bar u_{i,i+\textbf{a}} \phi _{i\alpha}^ + \phi _{i+\textbf{a}\alpha}  + u_{i,i+\textbf{a}} \phi _{i+\textbf{a}\alpha}^ +  \phi _{i\alpha} } \right]} \eea where $\bar u_{i,i+\textbf{a}}\, ( u_{i,i+\textbf{a}})$ are Hartree-Fock parameters to be determined self-consistently. We are interested in real parameters which do not depend on the lattice sites, but depend on the space directions $u_{i,i+\textbf{a}}=\bar u_{i,i+\textbf{a}}=u_{\textbf{a}}$. Then in momentum space representation, the Hamiltonian \eqref{JTH-F} has the form

\be\label{JTH-F2} H_{\rm H - F}  = \frac{{N}}{2}\sum\limits_{\textbf{a}}u_{\textbf{a}}^2J^{\textbf{a}}  - 2\lambda MN+ \sum\limits_k {{\varepsilon _k}\phi^+_k\phi_k },\ee where $N$ is the number of lattice sites and $\varepsilon_k$ is the dispersion of the $\phi_k$-boson (spinon) \be {\varepsilon _k} = \lambda  - {J_x}{u_x}\cos {k_x} - {J_y}{u_y}\cos {k_y} - {J_z}{u_z}\cos {k_z}\ee
The free energy of the theory with Hamiltonian \eqref{JTH-F2} is \be\label{JTfreeE}
F = \frac 12 \sum\limits_{\textbf{a}}u_{\textbf{a}}^2J^{\textbf{a}}   - 2\lambda M + \frac{{2T}}{N} \sum\limits_k {\ln \left( {1 - e^{ - \frac{{\varepsilon _k }}{T}} } \right)},\ee where $T$ is the temperature. The equations for the parameters $u_{\textbf{a}}$ and $\lambda$ are given by:
\be \frac{\partial F}{\partial u_{\textbf{a}}}=0 \qquad \qquad \frac {\partial F}{\partial \lambda}=0 \label{sysC}\ee

To ensure correct definition of the Bose theory \eqref{JTH-F2}, i.e. to have $\varepsilon_k \geq 0$ when the wave vector k runs over the first Brillouin zone of a cubic lattice, we have to make some assumptions for the parameter $\lambda$. For that purpose it is convenient to represent it in the form $\lambda  = \sum\limits_{\textbf{a}}\left(u_{\textbf{a}} J^{\textbf{a}}+ \mu u_{\textbf{a}}\right)$. In terms of the new parameter the $\phi_k$-boson dispersion is \be \varepsilon_k = \sum\limits_{\textbf{a}} \left[u_{\textbf{a}} J^{\textbf{a}} \left(1-\cos k_{\textbf{a}}\right)+\mu u_{\textbf{a}}\right]\ee and the theory is well defined for positive constants $u_{\textbf{a}}\geq0$ and $\mu \geq 0$.

We find the parameters $u_a$ and $\mu$ by solving the system \eqref{sysC}. For high enough temperatures both $\mu(T)$ and $u_{\textbf{a}}(T)$ are positive, and the excitation is gapped. Decreasing the temperature leads to decrease of $\mu(T)$. At temperature $T_C$ it becomes equal to zero $\mu(T_C)=0$, and long-range excitation emerges in the spectrum. Therefore the temperature at which $\mu$ reaches zero is the Curie temperature. We set $\mu=0$ in \eqref{sysC} and obtain a system of equations for the Curie temperature $T_C$ and $u_{\textbf{a}}$

\bea\label{TCurie}
& & \displaystyle u_{\textbf{a}'} = \frac{2}{N}\sum\limits_k \frac{\cos k_{\textbf{a}'}}{e^{\frac{1}{M T_C}\sum\limits_{\textbf{a}} u_{\textbf{a}} \rho^{\textbf{a}}(1-\cos k_{\textbf{a}})}  - 1}\nonumber\\
& & \displaystyle M = \frac{1}{N}\sum\limits_k \frac{1}{e^{\frac{1}{MT_C}\sum\limits_{\textbf{a}} u_{\textbf{a}} \rho^{\textbf{a}}(1-\cos k_{\textbf{a}})}  - 1} \eea

\begin{figure}[!t]
\includegraphics[width=.98\linewidth]{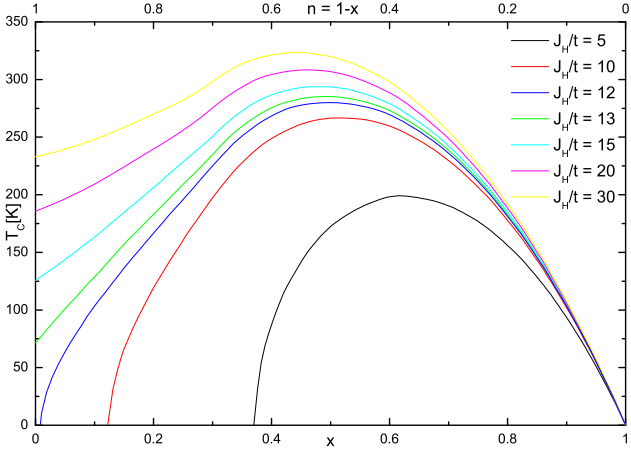}
\caption{(Color online) Curie temperature $T_C$ as a function of hope doping $x$ for different values of Hund's coupling $J_{\rm H}/t$, corresponding to the cases depicted in Fig. \ref{fig4}.}\label{fig3}
\end{figure}

The results for the Curie temperature $T_C$ as a function of doping $x$ are plotted in Figure \ref{fig3} for different values of the Hund's coupling in the absence of Jahn-Teller distortions and antiferromagnetism. One can see that with the increase of $J_{\rm H}/t$ the Curie temperature also increases and the maximum of the curves is shifted to lower values of $x$. The ends of the curves correspond to the transition from ferromagnetism to antiferromagnetism and for high enough value of $J_{\rm H}/t$ we have ferromagnetism across all values of $x$. This behavior is closely related to the behavior of the spin-stiffness curves (see figure \ref{fig4}).

\section{Critical temperature in the A-type antiferromagnetic regime}

Let us now turn to calculating the critical temperature in the A-type antiferromanetic phase. This phase is characterized by two positive and one negative effective exchange constants, namely $J_x >0$, $J_z >0$ and $J_y <0$. The Hamiltonian then reads:

\bea h = \sum\limits_i\big[ - {J_x}{{\bf M}_i} \cdot {{\bf M}_{i + {e_x}}} + |{J_y}|{{\bf M}_i} \cdot {{\bf M}_{i + {e_y}}}\nonumber\\
- {J_z}{{\bf M}_i} \cdot {{\bf M}_{i + {e_z}}} \big] \eea

We rewrite the vectors ${\bf  M}_i$ in terms of Schwinger bosons and use relation \eqref{eqM} for the $x$ and $z$ component, while for the $y$ component we use:

\bea\hskip -.5cm \textbf{M}_i \cdot \textbf{M}_j  = -\frac{1}{2}\left( {\phi_{i1}^+\phi_{j2}^+-\phi_{i2}^+\phi_{j1}^+} \right)\left( {\phi_{i1}\phi_{j2}-\phi_{i2}\phi_{j1}} \right)\nonumber\\
+\frac{1}{4}\left( {\phi _{i\alpha}^ +  \phi _{i\alpha} }\right)\left( {\phi _{j\beta}^ +  \phi _{j\beta} }\right)\eea

The second term here is a constant and we omit it. Introducing a term to ensure the Schwinger bosons constraint, we rewrite the Hamiltonian as
\bea H =  - \frac{1}{2}\sum\limits_i \Big[ {J_x}\left( {\varphi _{i\sigma }^ + {\varphi _{i + {e_x}\sigma }}} \right)\left( {\varphi _{i + {e_x}\sigma '}^ + {\varphi _{i\sigma '}}} \right) +\nonumber\\
|{J_y}|\left( {\varphi _{i1}^ + \varphi _{i + {e_y}2}^ +  - \varphi _{i2}^ + \varphi _{i + {e_y}1}^ + } \right)\left( {{\varphi _{i1}}{\varphi _{i + {e_y}2}} - {\varphi _{i2}}{\varphi _{i + {e_y}1}}} \right) \nonumber +\\
{J_z}\left( {\varphi _{i\sigma }^ + {\varphi _{i + {e_z}\sigma }}} \right)\left( {\varphi _{i + {e_z}\sigma '}^ + {\varphi _{i\sigma '}}} \right) \Big]  +\nonumber\\
\lambda\sum\limits_i {\left( {\varphi _{i\sigma }^ + {\varphi _{i\sigma }} - 2s} \right)} \hspace*{1cm}\eea

As with the ferromagnetic phase, we treat the four-boson interaction in Hartree-Fock approximation, with the effective Hamiltonian reading:

\begin{widetext}\bea {H_{\rm HF}} = \frac{1}{2}\sum\limits_i \bigg[ {J_x}\left\langle {\varphi _{i\sigma }^ + {\varphi _{i + {e_x}\sigma }}} \right\rangle \left\langle {\varphi _{i + {e_x}\sigma '}^ + {\varphi _{i\sigma '}}} \right\rangle +|{J_y}|\left\langle {\varphi _{i1}^ + \varphi _{i + {e_y}2}^ +  - \varphi _{i2}^ + \varphi _{i + {e_y}1}^ + } \right\rangle \left\langle {{\varphi _{i1}}{\varphi _{i + {e_y}2}} - {\varphi _{i2}}{\varphi _{i + {e_y}1}}} \right\rangle  \nonumber\\
+{J_z}\left\langle {\varphi _{i\sigma }^ + {\varphi _{i + {e_z}\sigma }}} \right\rangle \left\langle {\varphi _{i + {e_z}\sigma '}^ + {\varphi _{i\sigma '}}} \right\rangle - {J_x}\Big( {\left\langle {\varphi _{i\sigma '}^ + {\varphi _{i + {e_x}\sigma '}}}\! \right\rangle \varphi _{i + {e_x}\sigma }^ + {\varphi _{i\sigma }} + \left\langle {\varphi _{i + {e_x}\sigma '}^ + {\varphi _{i\sigma '}}} \!\right\rangle \varphi _{i\sigma}^ + {\varphi _{i + {e_x}\sigma}}} \Big) \nonumber\\
-|{J_y}|\Big(\!\!\left( {\varphi _{i1}^ + \varphi _{i + {e_y}2}^ +  - \varphi _{i2}^ + \varphi _{i + {e_y}1}^ + } \right)\!\left\langle {{\varphi _{i1}}{\varphi _{i + {e_y}2}} - {\varphi _{i2}}{\varphi _{i + {e_y}1}}} \right\rangle +\left( {{\varphi _{i1}}{\varphi _{i + {e_y}2}} - {\varphi _{i2}}{\varphi _{i + {e_y}1}}} \right)\left\langle {\varphi _{i1}^ + \varphi _{i + {e_y}2}^ +  - \varphi _{i2}^ + \varphi _{i + {e_y}1}^ + } \right\rangle  \Big)- \nonumber\\
{J_z}\Big( \left\langle {\varphi _{i\sigma '}^ + {\varphi _{i + {e_z}\sigma '}}}\! \right\rangle \varphi _{i + {e_z}\sigma }^ + {\varphi _{i\sigma }} + \left\langle {\varphi _{i + {e_z}\sigma '}^ + {\varphi _{i\sigma '}}} \right\rangle \varphi _{i\sigma }^ + {\varphi _{i + {e_z}\sigma }} \Big) \bigg] +\lambda \sum\limits_i {\left(\varphi _{i\sigma }^ + {\varphi _{i\sigma }}  - 2s\right)}\hspace{1cm} \eea\end{widetext}

The Hartree-Fock parameters are given by

\be\left\langle {\varphi _{i\sigma }^ + {\varphi _{i + {e_\mu }\sigma }}} \right\rangle  = \left\langle {\varphi _{i + {e_\mu }\sigma }^ + {\varphi _{i\sigma }}} \right\rangle  = {u_\mu } \qquad \mu=x,z\ee

\bea\left\langle {\varphi _{i1}^ + \varphi _{i + {e_y}2}^+ - \varphi _{i2}^+\varphi _{i + {e_y}1}^ + }\right\rangle\hskip -1cm &  \nonumber\\
= &\left\langle {{\varphi _{i1}}{\varphi _{i + {e_y}2}} - {\varphi _{i2}}{\varphi _{i + {e_y}1}}} \right\rangle  = {u_y}\hskip 1cm\eea
and we have again chosen them to be real parameters which do not depend on the lattice site $i$, but depend on the lattice direction $a$. We then split the Hamiltonian into classical and quantum parts

\be{H_{\rm HF}} = {h_{\rm cl}} + {h_{\rm q}},\ee where the classical part is given by

\be {h_{\rm cl}} = \frac{{{J_x}}}{2}u_x^2N + \frac{{|{J_y}|}}{2}u_y^2N + \frac{{{J_z}}}{2}u_z^2N - 2\lambda sN, \ee
while for $h_{\rm q}$ we have

\bea {h_q} =-\sum\limits_i \big[\frac{{{J_x}}}{2}{u_x}\left( {\varphi _{i + {e_x}\sigma }^ + {\varphi _{i\sigma }} + \varphi _{i\sigma '}^ + {\varphi _{i + {e_x}\sigma '}}} \right) +\nonumber\\
\frac{{|{J_y}|}}{2}{u_y}\left( {\varphi _{i1}^ + \varphi _{i + {e_y}2}^ +  - \varphi _{i2}^ + \varphi _{i + {e_y}1}^ +  + {\varphi _{i1}}{\varphi _{i + {e_y}2}} - {\varphi _{i2}}{\varphi _{i + {e_y}1}}}\! \right) \nonumber\\
+\frac{{{J_z}}}{2}{u_z}\left( {\varphi _{i + {e_z}\sigma }^ + {\varphi _{i\sigma }} + \varphi _{i\sigma }^ + {\varphi _{i + {e_z}\sigma }}} \right) \Big]  + \lambda \sum\limits_i {{\varphi _{i\sigma }^ + {\varphi _{i\sigma }}}} \hspace{.8cm}\eea
This Hamiltonian can be rewritten as \be {h_{\rm q}} = \sum\limits_k {\left[ {{\varepsilon _k}\varphi _{k\sigma }^ + {\varphi _{k\sigma }} + {\gamma _k}\varphi _{1k}^ + \varphi _{2k}^ +  + \gamma _k^*{\varphi _{1k}}{\varphi _{2k}}} \right]} \ee where we have introduced

\be {\varepsilon _k} = \lambda  - {J_x}{u_x}\cos {k_x} - {J_z}{u_z}\cos {k_z}\label{disp}\ee and

\be {\gamma _k} =  - i{J_y}{u_y}\sin {k_y}.\ee

We can easily diagonalize $h_q$ to

\be {h_{\rm q}} = \sum\limits_k {\left[ {{E_k}f_{k\sigma }^ + {f_{k\sigma }} + E_k^0} \right]} \ee with

\be {E_k} = \sqrt {\varepsilon _k^2 - |{\gamma _k}{|^2}} \qquad E_k^0 = \sqrt {\varepsilon _k^2 - |{\gamma _k}{|^2}}  - {\varepsilon _k}\ee

The free energy of the system is then given by

\bea {\cal F} = \frac{{{J_x}}}{2}u_x^2 + \frac{{|{J_y}|}}{2}u_y^2 + \frac{{{J_z}}}{2}u_z^2 - 2\lambda s \nonumber\\
+ \frac{2}{{\beta N}}\sum\limits_k {\ln \left( {1 - {e^{ - \beta {E_k}}}} \right)}  + \frac{1}{N}\sum\limits_k {E_k^0}. \eea

We can now construct a system of four equations for the parameters $u_x$, $u_y$, $u_z$ and $\lambda$:

\be \frac{\partial F}{\partial u_{\textbf{a}}}=0 \qquad \qquad \frac {\partial F}{\partial \lambda}=0, \ee 

In explicit form the system reads: \be\begin{array}{l}\displaystyle
{u_x} = \frac{1}{N}\sum\limits_k {\frac{{{\varepsilon _k}\cos {k_x}}}{{{E_k}}}} \left( {1 + 2{n_k}} \right)\\[13pt]
\displaystyle 1 = \frac{{|{J_y}|}}{N}\sum\limits_k {\frac{{{{\sin }^2}{k_y}}}{{{E_k}}}} \left( {1 + 2{n_k}} \right)\\[13pt]
\displaystyle {u_z} = \frac{1}{N}\sum\limits_k {\frac{{{\varepsilon _k}\cos {k_z}}}{{{E_k}}}} \left( {1 + 2{n_k}} \right)\\[13pt]
\displaystyle 2s + 1 = \frac{1}{N}\sum\limits_k {\frac{{{\varepsilon _k}}}{{{E_k}}}} \left( {1 + 2{n_k}} \right)\end{array}\label{TNeel}\ee
where $n_k$ is the Bose occupation number \be{n_k} = \frac{1}{{{e^{{E_k}/T}} - 1}}\ee and $E_k$ is the dispersion \be {E_k} = \sqrt {{{\left( {\lambda  - {J_x}{u_x}\cos {k_x} - {J_z}{u_z}\cos {k_z}} \right)}^2} - |{\gamma _k}{|^2}} \ee

The critical temperature is obtained for $\lambda = J_x u_x + |J_y|u_y + J_z u_z$. Substituting into \eqref{disp} we obtain

\be \varepsilon_k = J_x u_x \left( 1 - \cos k_z \right) + J_z u_z \left( 1 - \cos k_z \right) + |J_y|u_y.\ee

The dispersion at the critical temperature $T=T_N$ then has the form \begin{widetext} \be {E_k} = \sqrt {\left(J_x u_x \left( 1 - \cos k_x \right) + J_z u_z \left( 1 - \cos k_z \right) + |J_y| u_y \right)^2 - J_y^2 u_y^2\sin^2 k_y} \ee and $E_{k_i}=0$ has two solutions for $k_1 = \left(0,\frac \pi 2,0\right)$  and $k_2 = \left(0,-\frac \pi 2,0\right)$ respectively. Near the zero points, the dispersion adopts the form:

\be {E_{k \to {k_i}}} \approx \sqrt {{J_x}{u_x}|{J_y}|{u_y}k_x^2 + {J_z}{u_z}|{J_y}|{u_y}k_z^2 + J_y^2u_y^2{{\left( {{k_y} \mp \frac{\pi }{2}} \right)}^2}}\label{afmdisp} \ee
\end{widetext}

Thus, the magnons in the antiferromagnetic phase have dispersion that behaves as $\sim |\vec k|$ at small impulses.

Using the systems of equations for the Curie and N\'{e}el temperatures, \eqref{TCurie} and \eqref{TNeel} respectively, we can build the phase portraits for the $x<0.5$ region.

\section{Results}

In this section we will apply all the information presented so far to construct a realistic phase portrait of $\rm La_{1-x}Ca_xMnO_3$. As discussed in the introduction, we have to describe four phases, namely G-type antiferromagnet, C-type antiferromagnet, ferromagnet, and A-type antiferromagnet in order of decreasing $x$. However, since the most important transport effects, metal-insulator transition and colossal magnetoresistance effect are observed in the ferromagnetic part of the manganites' phase diagram, we will focus only on the $x<0.5$ region.

Before introducing our results, let us first discuss the criteria we have used to construct the curves for the critical temperatures. Our goal is to be in both qualitative and quantitative agreement with the experimental results, and so we have aimed to reproduce curves with the following characteristics: a) the maximum of the ferromagnetic part is observed around $x=0.38$; b) the maximal value of $T_C$ is around 265 K; c) left and right of the maximum the curves are as close to the experimental ones as possible; d) transition to A-type antiferromagnetic phase occurs at $x=0.08$.

To achieve this, our first step is to determine appropriate value for $J_{\rm H}/t$. As we discussed in section II, values lower than $J_{\rm H}/t = 10$ cannot reproduce the observed physics. Since we also have to consider antiferromagnetic term and Jahn-Teller distortions, both of which suppress the ferromagnetic phase, we have chosen to work with two different values of $J_{\rm H}/t$, namely $J_{\rm H}/t = 15$ and $J_{\rm H}/t = 20$. This in turn allows us to establish a lower limit upon the value of the hopping parameter $t$, needed to reproduce the observed critical temperatures. For $J_{\rm H}/t = 15$ this value is $t=0.215$ eV, and for $J_{\rm H}/t = 20$ it is $t=0.20$ eV.

Once we have selected $J_{\rm H}/t$, we turn to the other parameters. We have to consider the antiferromagnetic exchange constants $J_{\rm AF}^x$ ($=J_{\rm AF}^y$) and $J_{\rm AF}^z$ and the Jahn-Teller energy $E_{\rm JT}=g^2/(2k)$. For the latter, it is generally agreed that its effect decreases with increase of the doping level $x$. For this reason we choose to work with distortion, which splits the exchange constants $J_x$ and $J_y$ at $x=0.20$ and leads to the phase transition at $x=0.08$ (see fig.\;\ref{fig5}). This is important, since if the distortion is not included, we cannot explain the existence of the A-type antiferromagnetic phase. In addition, the absence of Jahn-Teller distortion above $x=0.20$ means that for fixed values of $J_{\rm H}/t$ the slope of $T_C[x]$ curve is controlled by $J_{\rm AF}^x$ and $J_{\rm AF}^z$.

\begin{figure}[!b]
\includegraphics[width=.97\linewidth]{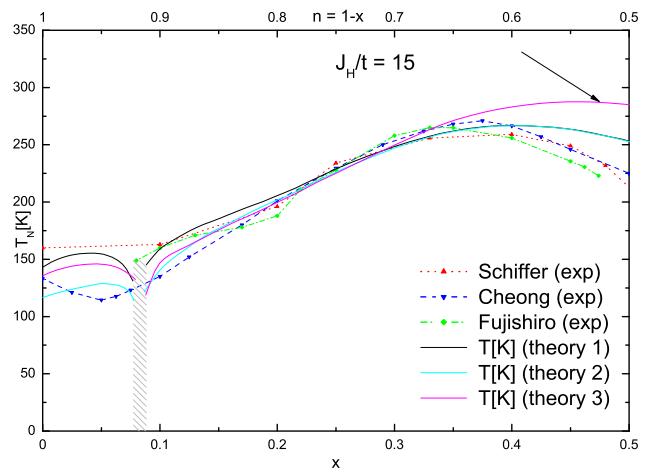}
\caption{(Color online) Critical temperature as a function of hole doping for $J_{\rm H}/t=15$ and: $t=0.365$ eV, $J_{\rm AF}^z =0.0140$, eV $J_{\rm AF}^x =0.000312$ eV (black line); $t=0.37$ eV, $J_{\rm AF}^z =0.0142$ eV, $J_{\rm AF}^x =0.000421$ eV (cyan line); $t=0.37$ eV, $J_{\rm AF}^z = 0.00632$ eV, $J_{\rm AF}^x =0.000316$ eV (magenta line).}\label{fig6}
\end{figure}

To better explain the impact of the antiferromagnetic constants on the Curie temperature curves, we have depicted in Figure \ref{fig6} the phase portraits for three different sets of parameters and fixed $J_{\rm H}/t=15$. The blue\cite{Cheong} (dashed), green\cite{Fuji1,Fuji2} (dash-dotted) and red\cite{Schiffer} (dotted) lines correspond to the experimental results, while the solid lines are obtained using the theoretical calculations presented here. The magenta line corresponds to the following set of parameters: $t=0.37$ eV, $J_{\rm AF}^z = 0.00632$ eV, and $J_{\rm AF}^x =0.000316$ eV.
The black line corresponds to $t=0.365$ eV, $J_{\rm AF}^z =0.0140$, and eV $J_{\rm AF}^x =0.000312$ eV, and the cyan one to $t=0.37$ eV, $J_{\rm AF}^z =0.0142$ eV, and $J_{\rm AF}^x =0.000421$ eV. With increase of $J_{\rm AF}^z$, the right end of the curve is shifted to lower values of $T_C$ and the maximum of the curve approaches the experimental value. Decreasing of $J_{\rm AF}^z$ has the opposite effect, and if the value drops below 0.006 eV, the maximum of the curve is no longer in the $x<0.5$ region. Therefore, larger values of $J_{\rm AF}^z$ are in better agreement with the experiment.

After we have chosen a value of $J_{\rm AF}^z$, which ensures that the right end and the maximum of our curves are close to the experimental ones, we examine the effect of $J_{\rm AF}^y$. Its value is important, since, together with the Jahn-Teller distortion, it controls the point at which $\rho_y$ reaches zero and the phase transition to A-type antiferromagnetism occurs. In order to ensure this transition happens at $x=0.08$, we have to either fix the value of $J_{\rm AF}^y$ and determine the needed $E_{\rm JT}$, or vice versa. With fixed value of $J_{\rm AF}^z$, $J_{\rm AF}^y$ also controls the slope of the curve left of the maximum, up to the point of phase transition. Increasing the value of $J_{\rm AF}^y$ lowers the effective constant $J_y$ and in turn lowers the value of the critical temperature, with the effect increasing when we approach $x=0.08$.

The right end on the curves in Figure \ref{fig6} however is not very close to the experimental ones. To bring the curves closer, one should further increase the value of $J_{\rm AF}^z$. However, high values of the antiferromagnetic constants suppress ferromagnetism, so in order to further increase $J_{\rm AF}^z$, we should first increase the value of $J_{\rm H}/t$ to make sure ferromagnetism persists up to $x=0.08$.

\begin{figure}[!ht]
\includegraphics[width=.99\linewidth]{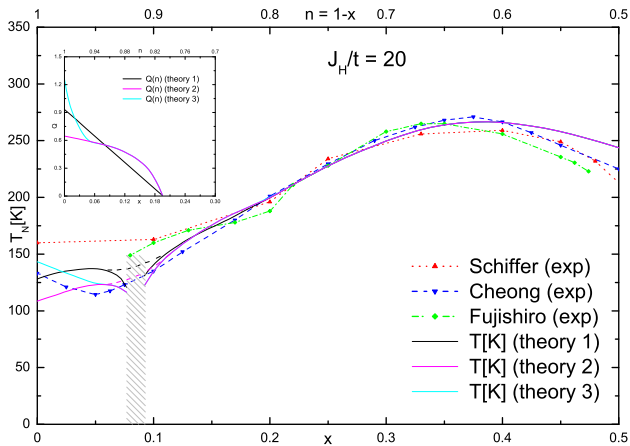}
\caption{(Color online) Critical temperature as a function of hole doping for $J_{\rm H}/t=20$, $t=0.462$ eV, $J_{\rm AF}^z = 0.0206$ eV, $J_{\rm AF}^x =0.00176$ eV, and different types of Jahn-Teller distortion. Inset: Jahn-Teller distortion as a function of carrier density (hoping).}\label{fig8}
\end{figure}

Following the procedure described in the beginning of this section, we have obtained the curves for the critical temperatures for $J_{\rm H}/t=20$ (Figure \ref{fig8}). The larger value for the Hund's constant allows us to work with larger values for the antiferromagnetic ones, namely $J_{\rm AF}^z = 0.0206$ eV and $J_{\rm AF}^x =0.00176$ eV, which in turn results in curves that are in better agreement with the experiment. To better illustrate the effect of the last remaining parameter, the Jahn-Teller distortion, we have used the same set of parameters for all the curves. Thus they are identical up to the point where distortion effects set off (around $x=0.2$). As in figure \ref{fig6}, blue (dashed), green (dash-dotted) and red (dotted) curves correspond to the experimental results.

We have worked with three different types of distortion, which are shown in the inset of Figure \ref{fig8}. Their crossing point represents the value of $Q(n)$, which we need in order to have transition to A-type antiferromagnetism at $x=0.08$ (i.e. to have $J_y=0$). This value is determined by the choice of $J_{\rm H}/t$ and $J_{\rm AF}^y$, so in this case it is equal for all three curves. In Figure \ref{fig8}, the black curve is a reference curve that corresponds to distortion, which increases linearly from $x=0.2$ to $x=0$. The magenta curve corresponds to distortion $Q(n)$, obtained by minimizing the fermion part of the free energy\cite{Us3}. Both the black and the magenta curves however, while giving values for $T_N$ very close to the experimental ones, decrease when $x$ approaches zero. To have the same behavior as the experimentally observed curve, the distortion has to grow rapidly for small values of $x$. Such behavior might be explained if we consider anharmonic terms in the phonon Hamiltonian. The cyan curve in Figure \ref{fig8} corresponds to phenomenologically fitted distortion, such that the resulting curve for $T_N$ has the same behavior as the experimentally observed one. It only deviates from the magenta curve in the A-type antiferromagnetic phase.

One can see that near the critical doping value, our curves start to rapidly decrease. The reason behind this effect is the decreasing value of $J_y$. As its value approaches zero, we are in effect describing the 2D case. Since our method of calculation is in agreement with the Mermin-Wagner theorem\cite{M-W}, we correctly obtain zero temperature at the critical doping value $x=0.08$. However this is not the observed experimental behavior and is a result of the limitations of our method, which can be avoided if we consider next to nearest neighbor corrections to the effective Hamiltonian \eqref{JTeff2}. For this reason, in our phase portraits we have shaded the region around $x=0.08$, where nearest neighbor contributions result in diminishing value of $J_y$. If one accounts for next to nearest neighbor corrections, the curves will continue to smoothly decrease past the critical point, which is represented by the dashed continuation lines near the critical point in Figure \ref{fig8}.

\section{Summary}

Starting from the well known Double Exchange Model and supplementing it with antiferromagnetic and Jahn-Teller terms, we have derived effective Heisenberg-type model for a vector, which describes the local orientation of the total magnetization. We have then used Schwinger-bosons mean-field theory in Hartree-Fock approximation to calculate the critical temperatures in the ferromagnetic and antiferromagnetic regimes. This technique of calculation is in agreement with the Mermin-Wagner theorem. We have then shown that the combination of these two ingredients provides results for the critical temperatures, which are in very good agreement with the experimental results.

We have argued that, in order to explain all the observed phases, one has to consider values for the Hund's coupling as large as $J_{\rm H}/t=20$. Another key point is to use antiferromagnetic constants, which depend on the lattice direction. Indeed, best agreement with the experimental results is observed when the value of $J_{\rm AF}^z$ is much larger than $J_{\rm AF}^x$ and $J_{\rm AF}^y$. Based on the agreement with the experiment, in Table \ref{t1} we provide estimation on the values of the model parameters, which best describe the observed physics. We have summarized the ``best fit'' values for both cases we have examined, namely $J_{\rm H}/t =15$ and $J_{\rm H}/t =20$.

\begin{table}[!b]
\caption{Estimation for the model parameters in eV for $J_{\rm H}/t =15$ and $J_{\rm H}/t =20$.\\}
\begin{tabular}{l@{\hskip .5cm}cccc}\hline\hline
&$t$&$J_{\rm AF}^x$&$J_{\rm AF}^y$&$J_{\rm AF}^z$\\ \hline
$J_{\rm H}/t =15$&0.37&0.000421&0.000421&0.0142\\
$J_{\rm H}/t =20$&0.462&0.00176&0.00176&0.0206\\\hline\hline
\end{tabular}\label{t1}\end{table}

While the method we have used is in very good agreement with the experimental results, it can still be improved. The inclusion of next to nearest neighbor corrections is needed to avoid the rapid decrease of the critical temperature curves near the point of phase transition. Anharmonic terms in the phonon Hamiltonian should also be considered and will possibly provide better agreement with the experiment in the low $x$ limit. The importance of the tolerance factor and related structural details such as $\rm Mn-O-Mn$ bond is another thing we have not considered.



\begin{thebibliography}{5}
\bibitem{Us1} N. Karchev and V. Michev, J. Phys.: Condens. Matter {\bf 19}, 156212 (2007).
\bibitem{Us2} V. Michev and N. Karchev, Phys. Rev. {\bf B76}, 174412 (2007).
\bibitem{Us3} V. Michev and N. Karchev, Phys. Rev. {\bf B80}, 012403 (2009).
\bibitem{Jonker1} G. H. Jonker and J. H. Van Santen, Physica {\bf 16}, 337--349 (1950).\\
\bibitem{Jonker2} J. H. Van Santen and G.H. Jonker, Physica {\bf 16}, 599--600 (1950).\\
\bibitem{Jonker3} G. H. Jonker, Physica {\bf 22}, 707--722 (1956).
\bibitem{Wollan} E. O. Wollan and W. C. Koehler, Physical Review {\bf 100}, 545--563 (1955).
\bibitem{Zener1} C. Zener, Physical Review {\bf 81}, 440--444 (1951).\\
\bibitem{Zener2} C. Zener, Physical Review {\bf 82}, 403--405 (1951).
\bibitem{AH} P. W. Anderson and H. Hasegawa, Physical Review {\bf 100}, 675--681 (1955).
\bibitem{deGennes} P. -G. de Gennes, Physical Review {\bf 118}, 141--154 (1960).
\bibitem{Goodenough} J. B. Goodenough, Physical Review {\bf 100}, 564--573 (1955).
\bibitem{Millis1} A. J. Millis, P. B. Littlewood, and B. I. Shraiman, Phys. Rev. Lett. {\bf 74}, 5144 (1995).
\bibitem{Roder} H. R\"oder and R. R. P. Singh and J. Zang, Phys. Rev. {\bf B56}, 5084 (1997).
\bibitem{Zang} J. Zang, A. R. Bishop and H. R\"oder, Phys. Rev. {\bf B53}, R8840 (1996).
\bibitem{Yunoki} S. Yunoki, A. Moreo, and E. Dagotto, Phys. Rev. Lett., {\bf 81}, 5612 (1998).
\bibitem{Sarma} D. Sarma, N. Shanthi, S. Barman, N. Hamada, H. Sawada, and K. Terakura, Phys. Rev. Lett. {\bf 75}, 1126 (1995).
\bibitem{Pickett1} W. E. Pickett and D. J. Singh, Europhys. Lett. {\bf 32}, 759 (1995).\\
\bibitem{Pickett2} W. E. Pickett and D. J. Singh, Phys. Rev. {\bf B53}, 1146 (1996).
\bibitem{Solovyev} I. Solovyev, N. Hamada, and K. Terakura, Phys. Rev. Lett. {\bf 76}, 4825 (1996).
\bibitem{Hotta} T. Hotta, Phys. Rev. {\bf B67}, 104428 (2003).
\bibitem{Millis2} A. J. Millis, R. Mueller, and B. I. Shraiman, Phys. Rev. {\bf B54}, 5389 (1996).\\
\bibitem{Millis3} A. J. Millis, R. Mueller, and B. I. Shraiman, Phys. Rev. {\bf B54}, 5405 (1996).
\bibitem{Held} Y. -F. Yang and K. Held, cond-mat/0903.2989, (2009).
\bibitem{Popovic} Z. Popovic and S. Satpathy, Phys. Rev. Lett., {\bf 84}, 1603 (2000).
\bibitem{Nolting1} M. Stier and W. Nolting, Phys. Rev. {\bf B75}, 144409 (2007).\\
\bibitem{Nolting2} M. Stier and W. Nolting, Phys. Rev. {\bf B78}, 144425 (2008).
\bibitem{M-W} N. D. Mermin and H. Wagner, Phys. Rev. Lett. {\bf 17}, 1133 (1966).
\bibitem{Pekker} D. Pekker, S. Mukhopadhyay, N. Trivedi, and P. M. Goldbart, Phys. Rev. {\bf B72}, 075118 (2005).
\bibitem{Golosov} D. I. Golosov, Phys. Rev. {\bf B71}, 014428 (2005).
\bibitem{Cheong} S.-W. Cheong and H. Y. Hwang, in Colossal Magnet oresistance Oxides, ed. Y. Tokura (1999).
\bibitem{Fuji1} H. Fujishiro, T. Fukase and M. Ikebe, J. Phys. Soc. Jpn. {\bf 70} 628 (2001).
\bibitem{Fuji2} H. Fujishiro and M. Ikebe, in Physics in Local Lattice Distortion, ed. H. Oyanagi and A. Bianconi, p. 433 (2001).
\bibitem{Kawano} H. Kawano, R. Kajimoto, M. Kubota, and H. Yoshizawa, Phys. Rev. {\bf B53}, R14709–R14712 (1996).
\bibitem{S-b1} D. P. Arovas and A. Auerbach, Phys. Rev. {\bf B38}, 316 (1988).
\bibitem{Schiffer} P. Schiffer,A. P. Ramirez, W. Bao, and S-W. Cheong,  Phys. Rev. Lett. {\bf 75}, 3336 (1995).
\end{thebibliography}
\end{document}